\documentclass[a4paper,english,nofootinbib,showpacs,preprint,aps]{revtex4}
\usepackage{color}
\usepackage{amssymb}
\usepackage{amsmath}
\usepackage{graphicx}
\usepackage[T1]{fontenc}
\usepackage{hyperref}
\usepackage[latin1]{inputenc}
\usepackage{amsfonts}

\begin{document}

\title{Induced fractional valley number in graphene with topological defects}
\author{Angel E. Obispo}
\email{ovasquez@feg.unesp.br}
\author{Marcelo Hott}
\email{marcelo.hott@pq.cnpq.br}
\affiliation{UNESP Universidade Estadual Paulista - Campus de Guaratinguetá - DFQ.\\
12516-410, Guaratinguetá-SP, Brazil.}

\begin{abstract}
We report on the possibility of valley number fractionalization in graphene
with a topological defect that is accounted for in Dirac equation by a
pseudomagnetic field. The valley number fractionalization is attributable to
an imbalance on the number of one particle states in one of the two Dirac
points with respect to the other and it is related to the flux of the
pseudomagnetic field. We also discuss the analog effect the topological
defect might lead in the induced spin polarization of the charge carriers in
graphene.
\end{abstract}

\maketitle

\section{Introduction}

Approximately seven years ago it was demonstrated by Hou, Chamon and Mudry
\cite{HCM} that fermion charge fractionalization can take place in monolayer
graphene due to Kekulé distortions that are described by means of a complex
valued scalar field (Higgs field) coupled to the Dirac field which, in its
turn, describes the dynamics of massless charge carriers in the quantum
field theory obtained as the low-energy regime of a discrete model
(tight-binding Hamiltonian) of nearest-neighbor hopping on the
two-dimensional honeycomb lattice \cite{Wallace, Semenoff}. The mechanism
for charge fractionalization relies upon the existence of zero-energy
eigenstates bounded to the vortex associated to the Kekulé distortions that
also would open an energy gap at the (two) previously present degeneracy
points - two inequivalent Dirac points in the Brillouin zone where the
valence and the conduction bands intersect each other - of the tight-binding
Hamiltonian. While the so-called valley symmetry is preserved by such midgap
bound-states, the vorticity of the distortion, either $n\geq 1 $ or $n\leq -1
$ , determines which one of the two triangular sublattices supports $|n|$
zero modes. That is, the zero modes do not exhibit the also called
pseudospin (sublattice) symmetry presented in both the tight-binding and
Dirac Hamiltonians. The zero modes and their sole contribution to the
fermion charge fractionalization were shown to persist in a chiral gauge
theory for graphene proposed by Jackiw and Pi \cite{Jackiw-Pi} by adding an
axial-vector gauge potential that incorporates axial gauge symmetry ($%
U_{A}(1)$) to the field theoretical model in \cite{HCM}.

The chiral coupling of electrons with gauge field in \cite{Jackiw-Pi} was
motivated to give a dynamical origin of the Higgs field, which together with
the gauge potential, enters in a phenomenological model
(Landau-Ginzburg-Abrikosov-Nielsen-Olesen model), whose minimum energy
solutions exhibit vortex-like profiles, as those proposed to be realized in
graphene. It worths to mention that a chiral gauge field was introduced much
earlier in the Dirac equation, as an effective equation for charge carriers
in a single layer of Carbon atoms to describe frustration in fullerenes \cite%
{Gonzalez}. Later, Pachos, Stone and Temme \cite{Pachos} shown that in a
honeycomb lattice the modulated hopping strength, otherwise described by
means of a real scalar field, may become a modulated complex-valued field if
one considers that the hexagonal lattice has its topology altered, such that
the former planar system becomes spherical, as in fullerenes. To take into
account the modified topology at the theoretical level, one introduces a
chiral gauge field in the Dirac Hamiltonian that describes the low-energy
dynamics of charge carriers. In this way, \cite{Pachos} has provided a
physical realization of magnetic vortices with fractionalized charge, as
those proposed originally in \cite{HCM} and \cite{Jackiw-Pi} and also
analyzed in detail in \cite{Chamon-et-al} and in \cite{Herbut} by taking
into account the effects of external magnetic fields.

In a remarkable work \cite{Semenoff} dedicated to the role gauge fields
might play in the dynamics of charge carriers described by an effective
quantum field theory of Dirac fermions, Semenoff showed that parity anomaly,
a phenomenon that was shown to happen in Quantum Electrodynamics in 2+1
space-time dimensions \cite{JDT}, could be simulated in a planar honeycomb
lattice with two species of atoms. In that scenario there is a difference in
energies of electrons localized on the different atoms and that implies in a
parity violating \textquotedblleft mass term\textquotedblright\ to the field
theory effective Hamiltonian that is eventually responsible for the
anomalous current that, paraphrasing Semenoff, \textquotedblleft would
couple to an unphysical external field of abnormal parity and is therefore
not directly observable\textquotedblright . In other words, that was an
axial-vector gauge potential entering in scene, at least theoretically, in
condensed matter, specially in graphite monolayer. Nowadays, it has been
proposed that very intense pseudomagnetic fields can be induced by strain in
graphene nanobubbles \cite{Levy} and such pseudomagnetic fields might be
associated to the curl of an axial-vector gauge potential non-minimally
coupled to Dirac spinors describing the dynamics of electrons in strained
graphene.

Indeed, fictitious magnetic fields in graphene have been very fruitful both
theoretically and experimentally, as for example, in the modeling of
corrugations and elastic deformations and in the study of \ influences that
topological defects could lead in the electronic properties of graphene \cite%
{meyer}-\cite{Kim}; in the experimental realization of Landau levels in very
intense (up to $300T$) pseudomagnetic fields due to stress in strained
graphene \cite{Levy} and the possibility to observe, also by means of a
scanning-tunneling-microscopy, Aharonov-Bohm interferences due to local
deformations (microstresses) in graphene \cite{Juan}. A good review on gauge
fields in graphene can be found in \cite{Vozmediano}, where the introduction
at theoretical level of elastic deformations, topological defects and
curvature in the low-energy effective Hamiltonian for graphene is reviewed
and some of their effects in the electronic properties are discussed (see
also \cite{Cortijothesis} and \cite{Fujita}). Additionally, fictitious
magnetic fields have been proposed as possible valleys filters in strained
graphene \cite{Zhai}- \cite{Fujita2}; a valley filter allows the
transmission of a current associated to only one of the valleys, while
filters the other one, that is, it transmits a valley-polarized current and
constitutes a crucial mechanism in \textit{valleytronics }\cite{Rycerz}, as
much as spin-filtering mechanism is fundamental to spintronics.

Motivated by many of the works cited above, we have shown \cite{us} that
vector and axial-vector gauge potentials by themselves can bind zero-energy
electrons and that fractional charge may be induced even in the absence of
Kekulé distortions. In this vein, we have also discussed the relation of
such induced fractional charge to the parity anomaly which would be realized
in gapped graphene as proposed almost thirty years ago\ \cite{Semenoff} and
in gapped graphene whose parity symmetry breaking term is provided by the
Haldane energy \cite{Haldane}. In addition we also discussed the possible
fractionization of another induced quantum number, which we had called
chiral charge (number) in the presence of magnetic and pseudomagnetic fields
and had briefly shown the connection of the chiral charge to the
time-component of the abnormal current found in \cite{Semenoff}.

The chiral charge (number) seems to have an important physical meaning and
must be treated as a physical observable in graphene whenever the valley
symmetry is preserved at the quantum Hamiltonian level. It is more
appropriate to call it \textit{valley number}, instead\textit{. }Since the
valley symmetry is manifest in this system and there is no intervalley
scattering, there is a doubling of fermions for each energy value and it is
legitimate to assign to each one-particle state a label (index) to account
for the fermion doubling. \ By its turn, the valley number is defined here
as the net number of valley states, i.e. $N_{\mathrm{v}}=N_{+}-N_{-}$, where
$N_{+}$ is the total number of states (summed over all energy states) around
to one of the Dirac points, while\ $N_{-}~$is the total number of states
associated to the other Dirac point.

Here, we consider the sample of graphene in the presence of an external
uniform magnetic field and with a local topological defect of disclination
type, which is equivalent to a pseudomagnetic field typical of a very thin
and long solenoid, the very same magnetic field usually used to discuss the
Aharonov-Bohm (AB) effect \cite{Furtado}-\cite{Cort}. We show that the
induced valley number is due only to the zero-energy modes (zero modes) of
one-particle states and that there is an imbalance in the number of
zero-energy eigenstates associated to the two Dirac points; as a
consequence, the induced net valley number is shown to be given by $%
N_{v}=\pm 1/2(1-2\{\Phi /2\pi \})$, where $0<\{\Phi /2\pi \}<1$ is the
fractional part of the reduced pseudomagnetic flux, whereas the global sign $%
\pm $ arises from the freedom to choose the zero-energy particles either in
the conduction ($+$) or in the valence ($-$) band. Some interesting features
of this result are, the independence of the valley number on the applied
external magnetic field, the fractionalization (even irrational values) of
the valley number for $\{\Phi /2\pi \}\neq 1/2$, and its null result for $%
\{\Phi /2\pi \}=1/2$. Moreover, by extending the concept of valley-filtering
further, we could say that this is a case of valley-polarized vacuum due to
a partial valley-filtering, when some of the zero-energy localized states of
one of the valleys are filtered.

In the next section we discuss the one-particle states by solving the proper
Dirac equation we have in hands and comment on the relevant symmetries of
the problem. In the third section we present the second quantization of the
fermion field, comment on the manifest valley-symmetry in the context of
quantum field theory, calculate the induced valley number and show that one
may have quasiparticles carrying fractional \textit{valley charge}. We also
compute the induced electric charge in this context.

To show the similarity of that supposed partial valley-filtering to a
partial spin-filtering also in graphene, in the fourth section we consider
the very same configurations of magnetic and pseudomagnetic fields in the
effective Hamiltonian for low-energy electrons whenever the valley degree of
freedom is decoupled and the relevant degrees of freedom are the pseudospin
(associated to the two triangular sublattices) and the spin polarization.
There we do not take into account the antiferromagnetic order as it was
considered in \cite{Semenoff2}, because that would break the spin symmetry
at the Hamiltonian level, what would not give reliability to the spin
polarization of the field as a physical observable. We show that the
filtering of some zero-energy states of spin polarization implies into
quasiparticles carrying fractional, even irrational, spin polarization. The
fifth section is left to further comments on the results we have found and a
short analysis about the possible induction of fractional valley number in
strained and in \textit{in-plane} deformed graphene samples.

\section{Bound states on magnetic and pseudomagnetic fields}

The effective field theory Hamiltonian describing the dynamics of the
electrons on the graphene honeycomb structure in the presence of an external
magnetic field $\overrightarrow{B_{V}}=\overrightarrow{\nabla }\times
\overrightarrow{V}$, and a pseudomagnetic field $\overrightarrow{B}_{A}=%
\overrightarrow{\nabla }\times \overrightarrow{A}$ can be written as
\begin{equation}
\mathcal{H}=\int d^{2}\vec{r}~\Psi ^{\dag }(\vec{r},t)\hat{h}_{D}\Psi (\vec{r%
},t),  \label{H}
\end{equation}%
where $\hat{h}_{D}=\overset{\rightarrow }{\alpha }.\left( -i\vec{\nabla}-e%
\overrightarrow{V}-\gamma _{5}\overrightarrow{A}\right) $ is the Dirac
Hamiltonian operator and $\Psi (\vec{r})~$is a four-component spinor, whose
transpose is $\Psi ^{T}=(\psi _{+}^{b}~\psi _{+}^{a}~\psi _{-}^{a}~\psi
_{-}^{b}).$ The superscripts $a$ and $b$ in the spinor components designate
the triangular sublattices where the electrons are supported on, while the
subscripts $\pm $ stand for each one of the two inequivalent Dirac
points.The matrix structure in (\ref{H}) is made explicit by means of the
following matrices
\begin{eqnarray}
\beta &=&\left(
\begin{array}{cc}
0 & I \\
I & 0%
\end{array}%
\right) ,~\vec{\alpha}=\left(
\begin{array}{cc}
\overset{\rightarrow }{\sigma } & 0 \\
0 & -\overset{\rightarrow }{\sigma }%
\end{array}%
\right) ,~  \notag \\
\alpha ^{3} &=&\left(
\begin{array}{cc}
\sigma _{3} & 0 \\
0 & -\sigma _{3}%
\end{array}%
\right) ,~\gamma _{5}=-i\alpha ^{1}\alpha ^{2}\alpha ^{3}=\left(
\begin{array}{cc}
I & 0 \\
0 & -I%
\end{array}%
\right) ,  \label{2}
\end{eqnarray}%
where $I$ is the $2\times 2$ identity matrix and $\overset{\rightarrow }{%
\sigma }=(\sigma _{1},\sigma _{2})$ and $\sigma _{3}$ are the Pauli matrices
in the standard representation.

We notice that (\ref{H}) is invariant under local $U(1)\times U_{A}(1)$
gauge transformations $\Psi \rightarrow e^{i(e\vartheta (\vec{x})+\gamma
_{5}\omega (\vec{x}))}\Psi $,$~$ $\overrightarrow{V}\rightarrow
\overrightarrow{V}+\overrightarrow{\nabla }\vartheta (\vec{x})~$and $%
\overrightarrow{A}\rightarrow \overrightarrow{A}+\overrightarrow{\nabla }%
\omega (\vec{x})$. Moreover, because we are in $(2+1)$ dimensions, the
chiral anomaly is absent.

The corresponding time-independent Dirac equation%
\begin{equation}
\hat{h}_{D}\Psi (\vec{r},t)=E\Psi (\vec{r},t)  \label{a1}
\end{equation}%
can be decomposed in two independent Dirac equations, for the upper $\psi
_{+}^{T}=(\psi _{+}^{b}~\psi _{+}^{a})$ and lower $\psi _{-}^{T}=(\psi
_{-}^{a}~\psi _{-}^{b})$ components of the energy eigenfunctions, that are
associated to each one of the Dirac points in the Brillouin zone of the
honeycomb lattice:

\begin{equation}
\left[ \left( \pm \overrightarrow{\sigma }\right) .\overrightarrow{\Pi }%
_{\pm }\right] \psi _{\pm }(\vec{r},t)=E\psi _{\pm }(\vec{r},t).  \label{a3}
\end{equation}%
with~$\overrightarrow{\Pi }_{\pm }=-(i\vec{\nabla}+e\overrightarrow{V}\pm
\overrightarrow{A})$.

The bispinors $\Psi _{+}^{T}=(\psi _{+}^{T}~0)$ and $\Psi _{-}^{T}=(0~\psi
_{-}^{T})$ are also eigenstates of $\gamma _{5}$ that commutes with the
Dirac Hamiltonian operator, with eigenvalues $\pm 1$, respectively. This is
the manifestation of the valley symmetry at the quantum mechanics level.
Furthermore, once the Dirac Hamiltonian operator anticommutes with $\alpha
^{3}$ and with $\alpha ^{3}\gamma _{5}$, the energy spectrum is symmetric
around the zero-energy level. In other words, if there is a norm-preserving
operator (unitary operator~$\mathcal{C}$) such that $\{\hat{h}_{D},\mathcal{C%
}\}=0$, then for each positive-energy normalized eigenstate, $\Psi _{|E|}$,
there is one corresponding negative-energy normalized eigenstate $\Psi
_{-|E|}=\mathcal{C}\Psi _{|E|}$ and the zero modes are self-conjugate.

When the external magnetic field is uniform and one considers that the
graphene sheet is under disclination, the vector and axial-vector potentials
can be written in the symmetric gauges, respectively as:%
\begin{equation}
V^{i}=-\frac{B}{2}\varepsilon ^{ij}x^{j}\ \ ,\ \ A^{i}=-\frac{\Phi }{2\pi
r^{2}}\varepsilon ^{ij}x^{j},  \label{3}
\end{equation}%
where $\varepsilon ^{12}=-\varepsilon ^{21}=1$ and $\Phi =\int d^{2}\vec{r}%
~B_{A}$ is the flux of the pseudomagnetic field $B_{A}=\Phi \delta (r)/2\pi
r $ (in cylindrical coordinates), which exhibits the same profile of the
magnetic field used to describe the AB effect. Such effect involves a
charged particle in the presence of a background magnetic field concentrated
within a flux tube where the probability of the particle be found is zero.
Here it is the topological defect that is represented by such
(pseudo)magnetic field (see \cite{Furtado}-\cite{Cort})

The zero-energy normalized eigenfunctions associated to the electrons under
the above magnetic and pseudomagnetic fields are given by
\begin{eqnarray}
\Psi _{0,l,+}(\vec{r}) &=&\sqrt{\frac{\left\vert eB/2\right\vert ^{1+l-\frac{%
\Phi }{2\pi }}}{\pi \Gamma (1+l-\frac{\Phi }{2\pi })}}\left(
\begin{array}{c}
e^{il\theta } \\
0 \\
0 \\
0%
\end{array}%
\right) r^{l-\frac{\Phi }{2\pi }}e^{-\frac{\left\vert eB\right\vert }{4}%
r^{2}}~\ \mathrm{and}  \notag \\
\Psi _{0,k,-}(\vec{r}) &=&\sqrt{\frac{\left\vert eB/2\right\vert ^{1+k+\frac{%
\Phi }{2\pi }}}{\pi \Gamma (1+k+\frac{\Phi }{2\pi })}}\left(
\begin{array}{c}
0 \\
0 \\
e^{ik\theta } \\
0%
\end{array}%
\right) r^{k+\frac{\Phi }{2\pi }}e^{-\frac{\left\vert eB\right\vert }{4}%
r^{2}},  \label{11}
\end{eqnarray}%
for$\mathrm{\ }eB>0$, and by
\begin{eqnarray}
\Psi _{0,l,+}(\vec{r}) &=&\sqrt{\frac{\left\vert eB/2\right\vert ^{1+l-\frac{%
\Phi }{2\pi }}}{\pi \Gamma (1+l-\frac{\Phi }{2\pi })}}\left(
\begin{array}{c}
0 \\
e^{-il\theta } \\
0 \\
0%
\end{array}%
\right) r^{l-\frac{\Phi }{2\pi }}e^{-\frac{\left\vert eB\right\vert }{4}%
r^{2}}~\mathrm{and}\text{~}  \notag \\
\Psi _{0,k,-}(\vec{r}) &=&\sqrt{\frac{\left\vert eB/2\right\vert ^{1+k+\frac{%
\Phi }{2\pi }}}{\pi \Gamma (1+k+\frac{\Phi }{2\pi })}}\left(
\begin{array}{c}
0 \\
0 \\
0 \\
e^{-ik\theta }%
\end{array}%
\right) r^{k+\frac{\Phi }{2\pi }}e^{-\frac{\left\vert eB\right\vert }{4}%
r^{2}},  \label{12}
\end{eqnarray}%
for $eB<0$. In the above expressions, $\theta ~$is the angular variable in
cylindrical coordinates and $\,l,k~\in ~%
%TCIMACRO{\U{2124} }%
%BeginExpansion
\mathbb{Z}
%EndExpansion
$, with $l>\Phi /2\pi -1~$and $k>-\Phi /2\pi -1$, which are conditions to
get normalized states. From those conditions one can note that there are
infinite zero modes in both valleys, and that the pseudomagnetic field
causes the lower bound of $k~$to be less than that of $l$ in the case $\Phi
>0$.\ This lead us to conjecture\ that there are additional states of $\
\Psi _{0,k,-}~$with respect to $\Psi _{0,l,+}$, and vice-versa in the case $%
\Phi <0$.

Solutions of the Dirac equations as those in (\ref{a3}) for the scattering
of massive fermions (when a mass term $m\sigma ^{3}$ is added to the Dirac
Hamiltonian operator) by only an AB-like magnetic field were discussed in
several papers, for instance \cite{HagenI}-\cite{Villalba} and of massless
fermions in \cite{Sitenko}. In our case, the presence of the uniform
magnetic field change the energy spectrum of the particle from a continuous
spectrum to a discrete one. Our treatment to reach the excited states, $%
E\neq 0$, follows the one given in \cite{Gitman1, Gitman2}, where it is
discussed the wavefunctions of massive fermions in the presence of an
uniaxial magnetic field with two contributions a uniform magnetic field and
the AB-like magnetic field. Parenthetically we notice that each one of the
equations in (\ref{a3}) corresponds exactly to the equation solved in \cite%
{Gitman1, Gitman2}, except that we are interested in massless fermions and
that the ground states were obtained here through the first-order
differential equations (\ref{a3}), while the second-order differential
equations are used in \cite{Gitman1, Gitman2} to obtain all the eigenstates,
but the irregular ones at the origin. (For the sake of simplicity we use a
specific value of the adjoint extension parameter; any other choice does not
modify the main results presented in the next section).

The orthonormal eigenstates of positive energy, angular momentum and of $%
\gamma _{5}~$can be worked out straightforwardly and have the forms%
\begin{eqnarray}
\Psi _{\left\vert E_{+}\right\vert ,l,+}(\vec{r}) &=&\frac{e^{il\theta }}{%
\sqrt{2}}\left(
\begin{array}{c}
R_{n,\nu _{+}}(r) \\
iR_{n-1,\nu _{+}+1}(r) \\
0 \\
0%
\end{array}%
\right) ,  \notag \\
\Psi _{\left\vert E_{-}\right\vert ,k,+}(\vec{r}) &=&\frac{e^{ik\theta }}{%
\sqrt{2}}\left(
\begin{array}{c}
0 \\
0 \\
R_{n,\nu _{-}}(r) \\
-iR_{n-1,\nu _{-}+1}(r)%
\end{array}%
\right) ,  \label{a18}
\end{eqnarray}%
where $\nu _{+}=l-\Phi /2\pi $, $\nu _{-}=k+\Phi /2\pi $, and%
\begin{equation}
R_{n,\nu }(r)=\sqrt{\frac{(\left\vert eB\right\vert /2)^{1+\nu }n!}{\pi
\Gamma (n+\nu +1)}}r^{\nu }e^{-\frac{\left\vert eB\right\vert }{4}%
r^{2}}L_{n}^{\nu }(\gamma r^{2})  \label{a21}
\end{equation}%
are the Gauss Laguerre modes in cylindrical coordinates conveniently
constructed such that $\int \left\vert R_{n,\nu }(r)\right\vert ^{2}d^{2}%
\overrightarrow{r}=1$, and we are assuming that $R_{-1,\nu }(r)=0$ and $\nu
>1$.

The bound states solutions for $\nu _{\pm }<-1~$are given by%
\begin{eqnarray}
\Psi _{\left\vert E_{+}\right\vert ,l,+}(\vec{r}) &=&\frac{e^{il\theta }}{%
\sqrt{2}}\left(
\begin{array}{c}
R_{n,\left\vert \nu _{+}\right\vert }(r) \\
-iR_{n,\left\vert \nu _{+}\right\vert -1}(r) \\
0 \\
0%
\end{array}%
\right) ,  \notag \\
\Psi _{\left\vert E_{-}\right\vert ,k,-}(\vec{r}) &=&\frac{e^{ik\theta }}{%
\sqrt{2}}\left(
\begin{array}{c}
0 \\
0 \\
R_{n,\left\vert \nu _{-}\right\vert }(r) \\
iR_{n,\left\vert \nu _{-}\right\vert -1}(r)%
\end{array}%
\right) ,  \label{6}
\end{eqnarray}%
while the eigenstates of negative energy with $\nu _{\pm }>+1$ and $\nu
_{\pm }<-1$ can be obtained by applying the unitary operator $\alpha ^{3}$
to the eigenstates (\ref{a18}) and (\ref{6}), respectively.

The subscripts $\left\vert E_{\pm }\right\vert $ designate the energy
eigenvalues which are the Landau levels (LL)

\begin{equation}
\left\vert E_{\pm }\right\vert =\sqrt{2eBn},\text{ with\ }n=1,2,3,...\text{
\textrm{for}}\left\{
\begin{array}{l}
l\geqslant 2+\left[ \Phi /2\pi \right] \\
k\geqslant 1-\left[ \Phi /2\pi \right]%
\end{array}%
\right.  \label{7}
\end{equation}%
that are obtained from the regularity condition $\nu _{\pm }>+1$ while, from
$\nu _{\pm }<-1$ one obtains%
\begin{equation}
\left\vert E_{\pm }\right\vert =\sqrt{2eB\left( n-m\pm \Phi /2\pi \right) }%
\text{ }(n=0,1,2,...,~),  \label{8}
\end{equation}%
where $m$ stands for $~l\leqslant -1+\left[ \Phi /2\pi \right] $ in case of $%
\left\vert E_{+}\right\vert $ and for $k\leqslant -2-\left[ \Phi /2\pi %
\right] $ in case of $\left\vert E_{-}\right\vert $, with $\left[ \Phi /2\pi %
\right] >0~$representing the integer part of the reduced pseudomagnetic flux.

Only regular eigenfunctions have been considered so far. The irregular ones,
corresponding to $\nu _{+}=-\{\Phi /2\pi \},1-\{\Phi /2\pi \}$ and to $\nu
_{-}=\{\Phi /2\pi \},-1+\{\Phi /2\pi \}$, ($0<\{\Phi /2\pi \}<1$ is the
fractional part of the reduced pseudomagnetic flux) are obtained from the
first-order differential equations (\ref{a3}). We have found that the
irregular eigenstates $\Psi _{\left\vert E_{+}\right\vert ,1+\left[ \frac{%
\Phi }{2\pi }\right] ,+}$ and $\Psi _{\left\vert E_{-}\right\vert ,-\left[
\frac{\Phi }{2\pi }\right] ,-}$ have their corresponding energy eigenvalues
in the set (\ref{7}), whereas for $\Psi _{\left\vert E_{+}\right\vert ,\left[
\frac{\Phi }{2\pi }\right] ,+}$ and $\Psi _{\left\vert E_{-}\right\vert ,-1-%
\left[ \frac{\Phi }{2\pi }\right] ,-}$ the corresponding energy eigenvalues
belong to the set (\ref{8}).

Notice that the LL (including the zero-energy level) are degenerate with
respect to eigenstates of $\gamma _{5}$ and $\hat{L}=-i\partial _{\theta }$.
The zero-energy level, one has$~l\geq \left[ \Phi /2\pi \right] ~$and $k\geq
-1-\left[ \Phi /2\pi \right] $, and the LL one has $l\geqslant 1+\left[ \Phi
/2\pi \right] $ and $k\geqslant -\left[ \Phi /2\pi \right] $, due to the
irregular eigenstates. The energy levels in the set (\ref{8}) with$%
~l\leqslant \left[ \Phi /2\pi \right] $ and $k\leqslant -1-\left[ \Phi /2\pi %
\right] $ may be degenerate only for $\{\Phi /2\pi \}=1/2$. Moreover, one
only has a complete set of eigenstates if the zero-energy eigenstates with $%
~l=\left[ \Phi /2\pi \right] $ and $k=-1-\left[ \Phi /2\pi \right] $ are
dropped out from the set of eigenstates. This complete set of eigenstates
comprises the basis in which the fermion field operator will be built from
in the next section.

\section{Induced valley number}

In this section we focus in the computation of the induced valley number
that is defined here as the vacuum expectation value of the valley number
operator, namely%
\begin{equation}
\hat{N}_{\mathrm{v}}=\frac{1}{2}\int d^{2}\overrightarrow{r}~\left[ \hat{\Psi%
}^{\dag }(\overrightarrow{r},t),\gamma _{5}\hat{\Psi}(\overrightarrow{r},t)%
\right] ,  \label{9}
\end{equation}%
where $\hat{\Psi}^{\dag }(\overrightarrow{r},t)$ is the fermion field
operator that is expanded in the basis of the one-particle states presented
in the previous section in the following way%
\begin{eqnarray}
\widehat{\Psi }(\overrightarrow{r},t) &=&\sum_{l=\left[ \Phi /2\pi \right]
+1}^{+\infty }\mathbf{\tilde{c}}_{0,l,+}\Psi _{0,l,+}(\overrightarrow{r}%
)+\sum_{k=-\left[ \Phi /2\pi \right] }^{+\infty }\mathbf{\tilde{c}}%
_{0,k,-}\Psi _{0,k,-}(\vec{r})+  \notag \\
&&+\sum_{n=0}^{+\infty }\left( \sum_{l}\left( \mathbf{c}_{n,l,+}\Psi
_{\left\vert E_{+}\right\vert ,l,+}(\vec{r})e^{-i\left\vert E_{+}\right\vert
t}+\mathbf{d}_{n,l,+}^{\dagger }\alpha ^{3}\Psi _{\left\vert
E_{+}\right\vert ,l,+}(\vec{r})e^{+i\left\vert E_{+}\right\vert t}\right)
+\right.  \notag \\
&&\left. +\sum_{k}\left( \mathbf{c}_{n,k,-}\Psi _{\left\vert
E_{-}\right\vert ,k,-}(\vec{r})e^{-i\left\vert E_{-}\right\vert t}+\mathbf{d}%
_{n,k,-}^{\dagger }\alpha ^{3}\Psi _{\left\vert E_{-}\right\vert ,k,-}(\vec{r%
})e^{+i\left\vert E_{-}\right\vert t}\right) \right) .  \label{10}
\end{eqnarray}

In this expansion of the fermion field $\mathbf{c}_{n,m,\mathrm{v}}$ ($%
\mathbf{d}_{n,m,\mathrm{v}}^{\dag }$) is the absorption (creation) operator
of a particle of energy $\left\vert E_{\pm }\right\vert $, angular momentum $%
m=l,k$ and valley $\mathrm{v}=\pm $ in the conduction (valence) band. The
energies $\left\vert E_{\pm }\right\vert $ of the particles are given by
expressions (\ref{7}) and (\ref{8}) with the corresponding values $l$ and $k$
can assume. The particles of zero-energy are, by construction of the fermion
field, supposed to be in the conduction band and their absorption operators
are represented by $\mathbf{\tilde{c}}_{0,m,\mathrm{v}}$ . The creation $%
\mathbf{d}_{n,m,\mathrm{v}}^{\dag }$ ($\mathbf{c}_{n,m,\mathrm{v}}^{\dag }$)
and absorption $\mathbf{d}_{n,m,\mathrm{v}}$ ($\mathbf{c}_{n,m,\mathrm{v}}$)
operators of particles in valence (conduction) band obey the following
anticommutation relations
\begin{eqnarray}
&&\left. \left\{ \mathbf{c}_{n^{{}},m^{{}},\mathrm{v}}~,\mathbf{c}%
_{n^{\prime },m^{\prime },\mathrm{v}^{\prime }}^{\dagger }\right\} =\left\{
\mathbf{d}_{n^{{}},i},\mathbf{d}_{n^{\prime },i^{\prime }}^{\dagger
}\right\} =\delta _{nn^{\prime }}\delta _{mm^{\prime }}\delta _{\mathrm{vv}%
^{\prime }},\right.  \notag \\
&&\left. \left\{ \mathbf{\tilde{c}}_{0^{{}},m^{{}},\mathrm{v}}~,\mathbf{%
\tilde{c}}_{0,m^{\prime },\mathrm{v}^{\prime }}^{\dagger }\right\} =\delta
_{mm^{\prime }}\delta _{\mathrm{vv}^{\prime }},\right.  \label{a25}
\end{eqnarray}%
with all the others possible anticommutators vanishing. Then, one can show
that $\left\{ \widehat{\Psi }^{\dag }(\overrightarrow{r},t),\widehat{\Psi }(%
\overrightarrow{r}^{\prime },t)\right\} =\delta ^{2}(\overrightarrow{r}-%
\overrightarrow{r}^{\prime })$, since one has a complete set of
eigenfunctions as that developed in the previous section.

Then, the valley number operator can be expressed as $\hat{N}_{\mathrm{v}}=%
\hat{N}_{\mathrm{+}}-\hat{N}_{-}$, where%
\begin{equation}
\hat{N}_{\mathrm{+}}=\sum_{l=\left[ \Phi /2\pi \right] +1}^{+\infty }\left(
\mathbf{\tilde{c}}_{0,l,+}^{\dag }\mathbf{\tilde{c}}_{0,l,+}-\frac{1}{2}\int
d^{2}\overrightarrow{r}\left\vert \Psi _{0,l,+}\right\vert ^{2}\right)
+\sum_{n}\sum_{l=-\infty }^{+\infty }\left( \mathbf{c}_{n,l,+}^{\dag }%
\mathbf{c}_{n,l,+}^{{}}-\mathbf{d}_{n,l,+}^{\dag }\mathbf{d}%
_{n,l,+}^{{}}\right) ,  \label{11a}
\end{equation}

and
\begin{equation}
\hat{N}_{\mathrm{-}}=\sum_{k=-\left[ \Phi /2\pi \right] }^{+\infty }\left(
\mathbf{\tilde{c}}_{0,k,-}^{\dag }\mathbf{\tilde{c}}_{0,k,-}-\frac{1}{2}\int
d^{2}\overrightarrow{r}\left\vert \Psi _{0,k,-}\right\vert ^{2}\right)
+\sum_{n}\sum_{k=-\infty }^{+\infty }\left( \mathbf{c}_{n,k,-}^{\dag }%
\mathbf{c}_{n,k,-}^{{}}-\mathbf{d}_{n,k,-}^{\dag }\mathbf{d}%
_{n,k,-}^{{}}\right)  \label{11b}
\end{equation}%
are the fermion number operator associated to the valley $+$ and valley $-$,
respectively. In obtaining these expressions one must remember that for each
positive-energy state of a given valley there is a corresponding
negative-energy state, such that they cancel each other. Furthermore, one
can note that each of the fermion number operators acquires a \textit{%
c-number }contribution\textit{\ }whose origin is due to the zero modes,
\textit{i.e.}%
\begin{eqnarray}
c-number_{+} &=&-\frac{1}{2}\int d^{2}\overrightarrow{r}~\sum_{l=\left[ \Phi
/2\pi \right] +1}^{+\infty }\left\vert \Psi _{0,l,+}\right\vert ^{2}=-\frac{1%
}{2}\int_{0}^{+\infty }d\zeta ~e^{-\zeta }\sum_{l=\left[ \Phi /2\pi \right]
+1}^{+\infty }\frac{\zeta ^{l-\Phi /2\pi }}{\Gamma \left( 1+l-q\Phi /2\pi
\right) }=  \notag \\
&=&-\frac{1}{2}\int_{0}^{+\infty }d\zeta ~\left( 1+\frac{\zeta ^{-\left\{
\Phi /2\pi \right\} }e^{-\zeta }}{\Gamma (1-\left\{ \Phi /2\pi \right\} )}-%
\frac{\Gamma (1-\left\{ \Phi /2\pi \right\} ,\zeta )}{\Gamma (1-\left\{ \Phi
/2\pi \right\} )}\right) =  \notag \\
&=&-\frac{1}{2}\left( \left\{ \Phi /2\pi \right\} +\int_{0}^{+\infty }d\zeta
\right) ,  \label{12a}
\end{eqnarray}

\begin{eqnarray}
c-number_{-} &=&-\frac{1}{2}\int d^{2}\overrightarrow{r}\sum_{k=-\left[ \Phi
/2\pi \right] }^{+\infty }\left\vert \Psi _{0,k,-}\right\vert ^{2}=-\frac{1}{%
2}\int_{0}^{+\infty }d\zeta ~e^{-\zeta }\sum_{k=-\left[ \Phi /2\pi \right]
}^{+\infty }\frac{\zeta ^{k+\Phi /2\pi }}{\Gamma \left( 1+k+\Phi /2\pi
\right) }=  \notag \\
&=&-\frac{1}{2}\int_{0}^{+\infty }d\zeta \left( 1+\frac{\zeta ^{\left\{ \Phi
/2\pi \right\} }e^{-\zeta }}{\Gamma (\left\{ \Phi /2\pi \right\} )}-\frac{%
\Gamma (\left\{ \Phi /2\pi \right\} ,\zeta )}{\Gamma (\left\{ \Phi /2\pi
\right\} )}\right) =  \notag \\
&=&-\frac{1}{2}\left( (1-\left\{ \Phi /2\pi \right\} )+\int_{0}^{+\infty
}d\zeta \right) ,  \label{12b}
\end{eqnarray}%
where $\Gamma (\alpha ,\zeta )$ is the Upper Incomplete Gamma Function \cite%
{Abram}, with $\zeta =eBr^{2}/2$. As a consequence each one of the fermion
numbers picks up a non-vanishing vacuum expectation, namely $N_{\mathrm{+}%
}=\left\langle 0\left\vert \hat{N}_{\mathrm{+}}\right\vert 0\right\rangle
=-\left( \frac{\Phi _{B}}{4\pi }+\left\{ \frac{\Phi }{4\pi }\right\} \right)
$ and $N_{\mathrm{-}}=\left\langle 0\left\vert \hat{N}_{\mathrm{-}%
}\right\vert 0\right\rangle =-\left( \frac{\Phi _{B}}{4\pi }+\frac{1}{2}%
-\left\{ \frac{\Phi }{4\pi }\right\} \right) $, where $\Phi _{B}/2\pi $ is
the reduced flux of the external magnetic field which is not quantized; it
is infinity in fact once $B_{V}$ is uniform and in the calculations we have
done so far the graphene sample has infinity dimensions ($r\in \lbrack
0,\infty )$). \ As one can see, there is an induced fermion charge
associated to each one of the valley due to the magnetic fields, whose
results are compatible with the results of the induced electric charge and
parity anomaly in QED in 2+1 dimensions. \ In fact, the total induced charge
of the system, which is also a physical observable and may be fractional
when specific configurations of the magnetic and pseudomagnetic fields are
considered and is fractional under Kekulé deformations - is
\begin{equation}
Q=\left\langle 0\left\vert \hat{Q}\right\vert 0\right\rangle =-\frac{e}{2}%
(-1+2e\frac{\Phi _{B}}{2\pi })  \label{13a}
\end{equation}%
From the above expression one sees that this induced charge is infinity once
the flux of the external and uniform magnetic field is infinity, but it also
has a fractional contribution $e/2$ due to the A-B-like pseudomagnetic field.

On the other hand the induced valley number is finite and given by
\begin{equation}
N_{\mathrm{v}}=\left\langle 0\left\vert \hat{N}_{\mathrm{v}}\right\vert
0\right\rangle =\left\langle 0\left\vert \hat{N}_{\mathrm{+}}-\hat{N}_{%
\mathrm{-}}\right\vert 0\right\rangle =\frac{1}{2}-\left\{ \frac{\Phi }{2\pi
}\right\} .  \label{13b}
\end{equation}%
That is a measure of the imbalance on the number of zero-modes among the
valleys. Parenthetically we notice that we would have $N_{\mathrm{v}%
}=-1/2(1-\left\{ \Phi /2\pi \right\} )$ had we considered the zero-energy
particles in the valence band.\qquad\ \ \ \ \ \

Another consequence of the \textit{c-number }as it stands in (\ref{12a}) and
in (\ref{12b}) is that each particle in the conduction or in the valence
band would carry a fractional valley charge as it follows
\begin{eqnarray*}
\hat{N}_{\mathrm{v}}\left\vert \left. 1_{\mathrm{cond}},n,m,\pm
\right\rangle \right. &=&\hat{N}_{\mathrm{v}}\mathbf{c}_{n,m,\pm }^{\dag
}\left\vert \left. 0\right\rangle \right. =(\pm 1+1/2(1-\left\{ \Phi /2\pi
\right\} ))\left\vert \left. 1_{\mathrm{cond}},n,m,\pm \right\rangle \right.
, \\
\hat{N}_{\mathrm{v}}\left\vert \left. 1_{\mathrm{val}},n,m,\pm \right\rangle
\right. &=&\hat{N}_{\mathrm{v}}\mathbf{d}_{n,m,\pm }^{\dag }\left\vert
\left. 0\right\rangle \right. =(\mp 1+1/2(1-\left\{ \Phi /2\pi \right\}
))\left\vert \left. 1_{\mathrm{val}},n,m,\pm \right\rangle \right. .
\end{eqnarray*}%
From this last result one can see that, except for the zero-energy states,
to each particle in the valence band and in the valley $+$ ($-$) is assigned
the same valley charge of a particle in the conduction band and in the
valley $-$ ($+$).

\section{Fractional (even irrational) spin polarization}

In this section we discuss the analogy between induced valley number and
induced spin polarization in graphene in order to reinforce the development
of valleytronics in analogy to spintronics. In valleytronics the
pseudomagnetic fields due to topological defects, microstresses and
deformations in clean graphene samples had shown to be useful in the
filtering mechanism which selects valley-polarized electric currents as much
as spin-polarized currents are produced in the context of spintronics.

In the previous two sections we have not considered the spin degrees of
freedom, since in clean samples of graphene there is no room to spin-spin
interaction or spin flipping in graphene. In the present section we decouple
the two valley degrees of freedom from the full Dirac Hamiltonian that
describes the low energy dynamics of electrons in the honeycomb lattice and
we are left with the pseudospin (sublattices $a$ and $b$) and spin degrees
of freedom. By following \cite{Semenoff2} we write the following quantum
mechanics Dirac Hamiltonian operator
\begin{equation}
\hat{h}=i\left( \overrightarrow{\sigma }\otimes I\right) .\left[
\overrightarrow{\bigtriangledown }-ie\overrightarrow{V}-i\left( I\otimes
\tau _{3}\right) \overrightarrow{A}\right] .  \label{16}
\end{equation}%
The $4\times 4$ matrix structure of $\hat{h}$ comes from the direct product
of $\ 2\times 2$ matrices, one of the sets, namely the identity matrix $I$
and the Pauli matrices $\sigma ^{i}$ are associated to the pseudospin, while
the other one is formed by the identity matrix and the spin Pauli matrices $%
\tau ^{a}$. Moreover, $\overrightarrow{V}$ is the vector gauge potential
associated to the external magnetic field and $\overrightarrow{A}$ is the
axial-vector gauge potential associated to the pseudomagnetic field. In
order to carry the analogy of induced valley number and induced spin
polarization further and show that the spin polarization can be fractional
as a consequence of the imbalance on the number of zero modes with different
spin polarizations, we consider that the gauge potentials have the same
configurations as given in (\ref{3}).

Because $\hat{h}$ commutes with the spin polarization operator $I\otimes
\tau _{3}$, one can label the energy eigenstates as $\Psi _{\pm }=\frac{1}{2}%
(1\pm \sigma _{3})\otimes I\Psi $, where $\Psi _{+}~$and $\Psi _{-}~$are
eigenstates of $I\otimes \tau _{3}$ also, with eigenvalues $+1$ and $-1$,$~$%
respectively. Moreover, $\hat{h}$ anticommutes with $\mathcal{C=}\sigma
_{3}\otimes \tau _{3}$, then if $\Psi _{\pm }^{E}~$is an eigenstate of $\hat{%
h}$ with eigenvalue $E\neq 0$, $\mathcal{C}\Psi _{\pm }^{E}~$is also an
eigenstate of $\hat{h}\ $with eigenvalue $-E$,\ and the\ zero modes are $%
\mathcal{C}$ self-conjugate. With such considerations one can find the
explicit form of the eigenstates of $\hat{h}$, $I\otimes \tau _{3}$ and $%
\hat{L}=-i\partial _{\theta }$ as they were found in section II. Moreover,
according to the steps developed in the previous section, only the zero
modes matters to the calculation of the vacuum expectation value. Then, for
the sake of simplicity we present only them here, namely%
\begin{eqnarray}
\Psi _{0,+,l} &=&\sqrt{\frac{(eB/2)^{1+l-\frac{\Phi }{2\pi }}}{\pi \Gamma
(1+l-\frac{\Phi }{2\pi })}}\left(
\begin{array}{c}
e^{il\theta } \\
0 \\
0 \\
0%
\end{array}%
\right) r^{l-\frac{q\Phi }{2\pi }}e^{-eBr^{2}/2}~\mathrm{and}\text{~}  \notag
\\
\Psi _{0,k,-}(\vec{r}) &=&\sqrt{\frac{(eB/2)^{1+k+\frac{\Phi }{2\pi }}}{\pi
\Gamma (1+k+\frac{\Phi }{2\pi })}}\left(
\begin{array}{c}
0 \\
e^{ik\theta } \\
0 \\
0%
\end{array}%
\right) r^{k+\frac{\Phi }{2\pi }}e^{-\frac{eB}{4}r^{2}}  \label{17}
\end{eqnarray}

We notice that the only thing that distinguishes these zero-energy
eigenstates from those in (\ref{11}) is their matrix structure. The same
happens to the other eigenstates. In other words, eigenstates $\Psi
_{\left\vert E\right\vert ,+,l}$ ($\Psi _{\left\vert E\right\vert ,-,k}$)
have in general the first and the third (the second and the fourth)
components different from zero. The energy eigenvalues are given by (\ref{7}%
) and (\ref{8}).

Since the quantum field operator
\begin{equation}
\hat{S}=\frac{1}{2}\int d^{2}\overrightarrow{r}~\left[ \hat{\Psi}^{\dag }(%
\overrightarrow{r},t),I\otimes \tau _{3}\hat{\Psi}^{\dag }(\overrightarrow{r}%
,t)\right]  \label{18a}
\end{equation}
commutes with the quantum field theory Hamiltonian operator%
\begin{equation}
\hat{H}=\frac{1}{2}\int d^{2}\overrightarrow{r}~\left[ \hat{\Psi}^{\dag }(%
\overrightarrow{r},t),i\left( \overrightarrow{\sigma }\otimes I\right) .%
\left[ \overrightarrow{\bigtriangledown }-ie\overrightarrow{V}-i\left(
I\otimes \tau _{3}\right) \overrightarrow{A}\right] \hat{\Psi}^{\dag }(%
\overrightarrow{r},t)\right]  \label{19}
\end{equation}%
it can be associated to a physical observable. Then, by following the same
steps of the section III one finds the induced spin polarization as given by%
\begin{equation}
S=\left\langle 0\left\vert \hat{S}\right\vert 0\right\rangle =\frac{1}{2}%
-\left\{ \frac{\Phi }{2\pi }\right\} .  \label{20}
\end{equation}%
One can check that the (quasi)particles would carry fractional (even
irrational) spin polarization. This result must be understood only as an
illustrative analogy to the result on the fractional valley number due to
topological defects in the graphene lattice, because topological defects do
not affect the spin polarization degree of freedom.

\section{Further comments and conclusions}

We have analyzed the influence of an external magnetic field in a clean
graphene sheet with a local topological defect (single disclination). Such a
topological defect can be described in the low-energy continuum model for
the charge carriers as a Aharonov-Bohm-like pseudomagnetic field whose
axial-vector gauge potential couples to the two valley degrees of freedom
with different signs. As a consequence of this and due to the configuration
of the pseudomagnetic field one has a partial breaking of the valley
symmetry originally present in the low-energy effective Hamiltonian for the
charge carriers. This partial breaking of the valley symmetry is revealed in
the energy spectrum of the system which is no longer the usual (degenerate)
Landau levels (LL) for a (relativistic) massless fermion in a uniform
magnetic field. For instance, between the zero-energy level and the first LL
with energy $\sqrt{2eB}\,$, there appears two additional non-degenerate
energy levels with energies $\sqrt{2eB\left\{ \Phi /2\pi \right\} }$ and $%
\sqrt{2eB(1-\left\{ \Phi /2\pi \right\} )}$ ($0<\left\{ \Phi /2\pi \right\}
<1$ is the fractional part of the reduced pseudomagnetic flux), each one of
these intermediary energy levels contains a unique member of only one of the
valleys. In addition, each one of the LL is partially degenerate as regards
the valley degree of freedom, namely there is an imbalance on the number of
valley states on those energy levels. \ A measure of this imbalance, which
we have called a partial filtering, can be defined in the context of quantum
field theory as the valley number given by the vacuum expectation value of
the number operator associated to one of the valleys minus the expectation
value of the number operator associated to the other valley. We have
analyzed that in the third section and have shown\ that due to the sole
contribution of the zero modes, the valley number is in fact finite and may
be fractional (even irrational) for $\left\{ \Phi /2\pi \right\} \neq 1/2$.
To show that the sum of states is infinity, specifically concerning the
zero-energy states, we have also computed the induced electric charge and
have shown that it is proportional do the flux of the external magnetic
field, which is infinity because the external field is uniform and the
graphene sample is taken to be infinity. This last result would be expected
from the parity anomaly in QED in 2+1 space-time dimensions. In fact, in the
absence of intervalley scattering, the continuum model for the dynamics of
charge carriers in graphene in magnetic and pseudomagnetic fields may be
seen as two decoupled QED in 2+1 dimensions (one for each valley degree of
freedom) for massless fermions.

The above splitting on the lowest LL may be a good candidate to explain the
degeneracy lift of the zero-energy level and, perhaps the appearance of a
mass gap, that has been observed in graphene samples under relatively strong
magnetic fields. Since it seems to be very difficult to have a perfect,
free-of-defects graphene sample, the influence of those kind of topological
defects enhanced by a relatively strong magnetic field might be observed.\

We have also wondered if similar effects could also take place in strained
samples of graphene under microstresses, as those reported in \cite{Levy}
where uniform (pseudo)magnetic fields up to $300T$ seems to be realized. In
view of the results in \cite{us}, we conclude that particles associated to
the different valleys would describe orbits with different cyclotron
frequencies, namely $\omega _{\pm }=\left\vert eB\pm B_{A}\right\vert /2$
and each Landau level, say $\sqrt{2eBn}$, is split in two levels with
energies $\sqrt{\left\vert eB+B_{A}\right\vert n}$\ and $\sqrt{\left\vert
eB-B_{A}\right\vert n}$, each one contains representative of only one of the
valleys. The zero-energy level still persists with representatives of both
valleys, but since they have different cyclotron frequencies, there will be
an induced valley number given by $N_{\mathrm{v}}=$\ $\pm \Phi /2\pi $\ (the
ambiguity of the sign is attributed to the zero-energy particles be assigned
to the valence or to the conduction band), which is no longer fractional,
neither is finite once the pseudomagnetic field is uniform and the sample is
taken to be infinitely large. An interesting aspect on these results is that
now we are able, by fine-tuning the external magnetic field, to filter one
of the valleys completely and leading to \textit{valley-polarized} cyclotron
orbits and to \textit{chiral eddy currents}.\

We have also analyzed other kinds of deformations in graphene, such as the
symmetric in-plane ones \cite{Juan}, but we have not been able to reach a
definite conclusion in favor of partial filtering in this case.
Nevertheless, in view of the results presented here and based on the index
theorems on the number of zero modes of the Dirac Hamiltonian \cite{Atiyah}-%
\cite{Jackiw}, we believe that a partial valley-filtering and a non-null
induced valley number take place whenever there is a net pseudomagnetic
flux.\ Then, for the example of in-plane deformation analyzed in \cite{Juan}%
, where the pseudomagnetic field is given by $B_{A}=\frac{-\kappa u_{00}\Phi
_{0}}{\pi \sigma ^{6}}(r^{4}-7\sigma ^{2}r^{2}+4\sigma
^{4})e^{-r^{2}/2\sigma ^{2}}$ we have found the pseudomagnetic flux as $\Phi
=4\kappa u_{00}\Phi _{0}$, and we would expect that, even in the background
of the uniform magnetic field for which the index theorem seems not to be
applicable since one has an infinite flux \cite{Niemi}, an induced valley
number given by $N_{\mathrm{v}}=$\ $\pm 2\kappa u_{00}\Phi _{0}$, where $\Phi _{0}$
is the flux quantum and $\kappa u_{00}=0.3$ from the data in \cite{Juan}.

\section{Acknowledgments}

AEO thanks to the CAPES/CNPq-IEL Nacional-Brasil program for the
scholarship. This work is also partially supported by CNPq (procs.
482043/2011-3, 306316/2012-9).

\end{document}